# Measuring times to determine positions

Amelia Carolina Sparavigna
Department of Applied Science and Technology
Politecnico di Torino, Italy

*Among the first devices used to measure the time we find the sundials and the water-clocks, as told by Vitruvius in his book on the Architecture. The sundials work because of the shadows cast by a rod or pole, the gnomon, on their basements. Besides being an instrument able to measure the time intervals, the sundial provided information on Earth and heaven to the ancient astronomers. Here we discuss how this device is able to determine the latitude and the north-south direction. The problem of the longitude is also shortly discussed.*

Among the first devices used to measure the time we find the sundials and the clepsydrae, that is, the water-clocks. This is what Marcus Vitruvius Pollio, who lived during the first century BC, tells us in his book on the Architecture [1,2]. The sundials work because of the shadows cast by a rod or pole, the gnomon, on their basements. Vitruvius is reporting that sundials had different forms. Those of semicircular form, he says, have been invented by Berosus the Chaldean; the "scaphes" or hemispheric sundials, by Aristarchus of Samos, the "arachne", by the astronomer Eudoxus, and so on. The Latin writer describes how to prepare the analemma of a sundial (Fig.1), following the shadow of the gnomon that he assumes having a vertical direction, that is the direction of the plumb-line. Besides being an instrument used to measure the time intervals, the Vitruvius' sundial can give us information of our position on the Earth.

**The equinoctial gnomon and the tilt of the Earth**
In the chapter of [2], entitled "The analemma and its applications", Vitruvius starts his discussion remembering the fact that "when the sun is at the equinoxes, that is, passing through Aries or Libra", the gnomon casts a shadow having a different length at different places. He is telling that the shadow is equal to 8/9 of the gnomon's length, latitude of Rome. If we use this ratio as the tangent of an angle, we find a value of it of about 41.63°. The latitude of Rome is 41.9° [3]. Vitruvius also reports other examples, he tells that in Athens, the shadow is equal to 3/4 of the gnomon's length, having then a latitude of 36.87° (37.96° [3]; at Rhodes to 5/7, that is 35.54°, (36.16° [3]); at Tarentum, to 9/11, that is 39.29° (40.47° [3]) and at Alexandria the ratio is equal to 3/5, which corresponds to an angle of 30.96 degrees (31.20° [3]). Therefore, wherever a sundial is to be constructed, "it is found that the shadows of equinoctial gnomons are naturally different from one another", because the latitude is different. Let us note that Vitruvius is naming the gnomon, the "equinoctial" gnomon.

At the noon on the Equinox, the ratio between the length of the shadow and the length of a gnomon, is the tangent of an angle corresponding to the latitude of the place. Let us note that the results of measurements are given by Vitruvius as fractions, according to the Greek mathematics. Moreover, the values are systematically lower than the actual ones.

Pliny in the chapter of his Natural History [4], entitled "Division of the Earth into parallels and shadows of equal length", tells that the Greeks invented the geographical divisions of the Earth, "in respect to the length of their days and nights, and in which of them the shadows are of equal length, and the distance from the pole is the same... The lines or segments which divide the world are many in number; by our people they are known as circles, by the Greeks they are called parallels". Let us remark that the ancient imagined the Earth as a sphere.

The Greeks determined the tropics, where at noon on the equinox there are no shadows, and the Arctic circles, having the complementary angles. Using the shadows of an instrument for measuring time, they had the latitude.

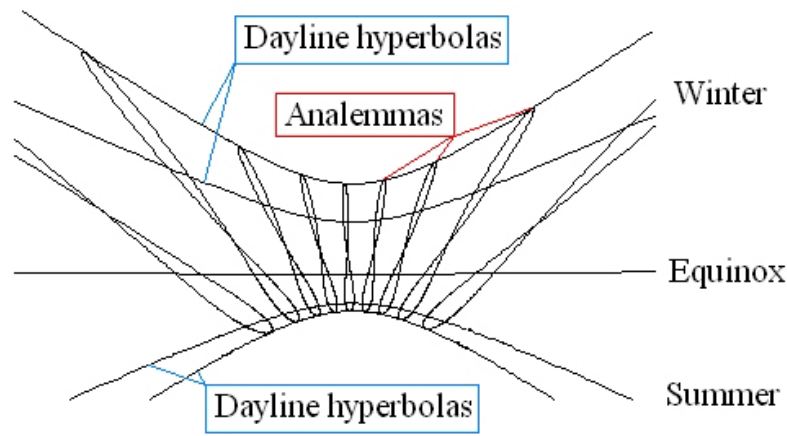

Fig.1 If we record the shadow positions for the same time each day we see that they form a figure looking like number 8. This is the analemma. The paths of the tip of the shadow for a single day are hyperbolas. This image is adapted from the figure by Richard D. Swensen, http://www.uwrf.edu/AboutUs/DesignOfTheRichardSwensenSundial.cfm

Of course, we have to wait the Equinox. What happens then at the Equinox at the main parallels?. Let us consider the Spring Equinox, approximately March 20-21. We have twelve hours of daylight and twelve hours of darkness at all points on the Earth's surface. Sunrise is at 6 a.m. and sunset is at 6 p.m. of the local solar time. The sun is on the horizon at the North Pole. At the Arctic Circle, the sun has a zenith angle of $66.5^{o}$, that is, it is low in the sky at $23.5^{o}$ above the horizon. At the Tropic of the Cancer, the sun has $23.5^{o}$ zenith angle. In other places, the sun is down from the zenith by an angle equal to their latitude.

The ancient town of Syene, the modern Aswan in Egypt, is approximately at the tropical latitude. The modern Aswan has a latitude of $24.08^{o}$. An equinoctial gnomon could give us a ratio of 0.447. According to Cleomedes, Eratosthenes used some measures obtained from sundials at Alexandria and at Syene, to have the circumference of the Earth [5]. And Vitruvius too is telling that he used the "equinoctial gnomons". As previously told, using them we can figure out the ratio of shadow and gnomon's length and have the latitude.

In a previous paper [5], I asked what Eratosthenes did with the equinoctial gnomons. In my opinion, he used them to evaluate the latitudes of Alexandria and Syene to have the Earth's radius. At the same time he also evaluated the inclination of ecliptic, that is the tilt of Earth's axis [6]. Supposing Syene at the tropical parallel, its latitude is this tilt. It is usually reported that Eratosthenes knew the fact that at noon on the Solstice the Sun is at the zenith of Syene, and that a gnomon cannot cast a shadow. It is quite sure however, that he also knew that at noon on the Equinox the gnomon cast a shadow giving the Syene latitude, and therefore the inclination of the Earth.

In several books it is reported that Eratosthenes determined the tilt of the Earth [7,8], but no details of the method he used are reported. Probably, it was simply the measure of the Syene latitudes. According to Ref.9, Ptolemy refined the Eratosthenes' measures, using the oscillation of the sundial shadows between solstices, obtaining a value of 11/83 of 180 degrees. The reference is also discussing the errors involved by measuring with sundials.

In the next section we will see what Vitruvius tells about Eratosthenes and the tilt of the heaven, in a chapter of his book concerning the urban planning.

**Due North**

The ancient Romans used a specific urban planning, developed for military castra, the military defence places. The basic plan consisted of a central forum surrounded by a rectilinear grid of streets. Two of them are slightly wider, intersecting in the middle to the central forum. One, the Decumanus, is running east–west, the other, the Cardo, is running north–south. Torino, Turin, has a well-preserved Roman urban scheme. The core of Turin is the Roman Quadrilateral, established in the first century BC, as the military Castra Taurinorum, that became Augusta Taurinorum, in honour of Augustus. We have streets at right angles, in the form of a square grid.

Cardo is the latin word for "hinge", therefore the cardo was the axis of the town, that one needs to know for the urban planning. Let us see what Vitruvius is telling in his chapter on "The directions of the streets with remarks on the winds" [2]. He discusses how to find the orientation of the cardo. "In the middle of the city place a marble amussium, laying it true by the level, or else let the spot be made so true by means of rule and level that no amussium is necessary. In the very centre of that spot set up a bronze gnomon ... At about the fifth hour in the morning, take the end of the shadow cast by this gnomon, and mark it with a point. Then, opening your compasses to this point which marks the length of the gnomon's shadow, describe a circle from the centre. In the afternoon watch the shadow of your gnomon as it lengthens, and when it once more touches the circumference of this circle and the shadow in the afternoon is equal in length to that of the morning, mark it with a point. From these two points describe with your compasses intersecting arcs, and through their intersection and the centre let a line be drawn to the circumference of the circle to give us the quarters of south and north". Again, using a measure of a time interval we have information on the direction due North.

In the same chapter, Vitruviius tells that "Eratosthenes of Cyrene, employing mathematical theories and geometrical methods, discovered from the course of the sun, the shadows cast by an equinoctial gnomon, and the inclination of the heaven that the circumference of the Earth is two hundred and fifty-two thousand stadia". Vitruvius is synthesizing perfectly what we have previously discussed on the Eratosthenes' measures of Earth and heaven.

**The problem of the longitude**
Before talking about the longitude, let me report the discussion written by Pliny the Elder in his chapter entitled "What regulates the daylight on the Earth" [4]. He tells the following: "In Africa and in Spain it is made evident by the Towers of Hannibal, and in Asia by the beacons, which, in consequence of their dread of pirates, the people erected for their protection; for it has been frequently observed, that the signals, which were lighted at the sixth hour of the day, were seen at the third hour of the night by those who were the most remote".
The problem of evaluating the longitude was quite old then, and, perhaps, Pliny had a faint idea on the use of time to get it. It is only a precise measurement of time, that can solve the problem of the longitude determination.
It seems that it was during the second century BC, that Hipparchus firstly proposed a system of determining the longitude by comparing the local time of a place with an absolute time [10]. Let us consider that Hipparchus discovered the precession of the equinoxes too, comparing his own determined ecliptic longitude (see Fig.2), which is the angular distance measured from the ecliptic to the first point of the Aries, of some stars from those of Timocharis, about 150 years earlier [11]. The first point, or cusp, is the imaginary line that separates a pair of consecutive signs in the zodiac. And the first point of Aries correspond to the equinox [12].

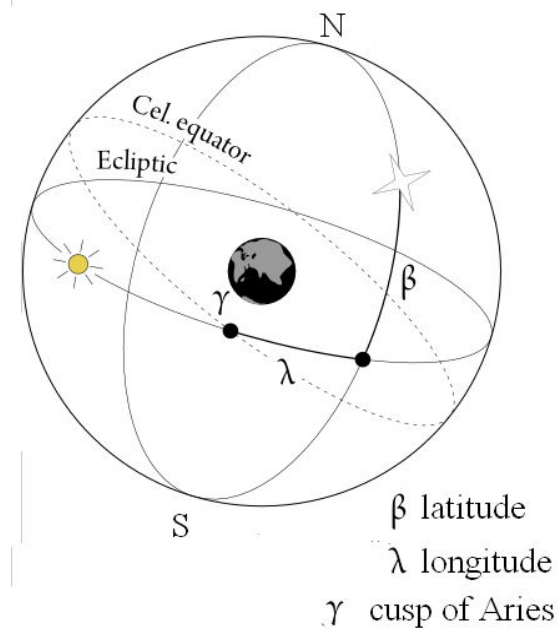

Fig.2 The ecliptic coordinate system (celestial), adapted from a figure by Joshua Cesa, Wikipedia.

The determination of the longitude was quite difficult for explores and navigators. Amerigo Vespucci was perhaps the first to propose a solution, after having "so much difficulty in determining it" [13]. He compared the relative positions of the Moon and Mars with their anticipated positions, to deduce his longitude. Of course, because he required a stable viewing platform for astronomical measures, his technique was useless on the rolling deck of a ship. In 1612, Galileo Galilei proposed to use the orbits of the Moons of Jupiter as a universal clock, but there was the same problem of the Vespicci's method, the rolling of the ship's deck. Without an universal clock, it is the use of the lunar distance, that is, the angle between the Moon and another celestial body, as proposed by Vespucci, that can help in determining the longitude. A navigator can use the measured lunar distance and a nautical almanac to have the Greenwich time [14].

The solution arrived eventually in 1773. John Harrison, an English clockmaker, invented the marine chronometer, allowing the determination of the longitude at sea [15], solving then a problem that lasted more than two thousands of years.

**References**
1. Marcus Vitruvius Pollio (born c.80–70 BC, died after c.15 BC) was a Roman architect and engineer, active during the 1st century BC. He is the author of the "De Architectura", that is "On Architecture". Gaius Plinius Secundus (23 AD – August 25, 79 AD), known as Pliny the Elder, was a Roman naturalist and naval commander of the Roman Empire. He wrote an encyclopaedic work, the "Naturalis Historia". Claudius Ptolemy (c.90 AD – c.168 AD), was a Roman citizen of Egypt who wrote in Greek. He was a mathematician, astronomer, geographer, astrologer, and poet of a single epigram in the Greek Anthology. He lived in the Roman Egypt, probably in Alexandria. Hipparchus, or Hipparchos (c.190 BC – c.120 BC), Greek astronomer and mathematician of the Hellenistic period, is considered the founder of trigonometry. He discovered the precesiion of equinoxes. Eratosthenes of Cyrene (c.276 BC – c.195 BC) was the Greek mathematician, geographer and astronomer, who measured the Earth circumference.
2. Marcus Vitruvius Pollio, The Architecture, Translated by Joseph Gwilt, Priestly and Weale, London, 1826; Vitruvius, The Ten Books on Architecture, Translated by Morris Hicky Morgan, Harvard University Press, 1914, http://www.gutenberg.org/files/20239/20239-h/29239-h.htm
3. For the latitude of the towns, I used the data of Wikipedia.
4. Pliny the Elder, Natural History, translated in English by John Bostock, http://www.perseus.tufts.edu/hopper/text?doc=Plin.+Nat.+toc
5. A.C. Sparavigna, Two amateurs astronomers at Berkeley, arXiv:1202.0950v1 [physics.pop-ph], 2012, http://arxiv.org/abs/1202.0950
6. The plane of the Earth's orbit is the ecliptic. The plane perpendicular to the rotation axis of the Earth is the equatorial plane. The angle between the ecliptic and the equatorial plane is the inclination of the ecliptic, or also the axial tilt of the Earth.
7. James Stuart Tanton, Encyclopedia of Mathematics, 2005, Facts On File.
8. J.H. Shirley and R.W. Fairbridge, Encyclopedia of planetary science,1997, Chapman & Hall

9. R.R. Newton, The sources of Eratosthenes' measurement of the Earth, Q. Jl R. astr. Soc. 21 (1980) 379-387

10. http://en.wikipedia.org/wiki/History_of_longitude

11. D.N. Mallik, Elements of astronomy, Cambridge University Press, 1931. In this reference, Timocharis appears as "Teniocharis". See also http://en.wikipedia.org/wiki/Timocharis

12. http://en.wikipedia.org/wiki/Ecliptic_longitude

13., http://en.wikipedia.org/wiki/Longitude

14. http://en.wikipedia.org/wiki/Lunar_distance_(navigation)

15. Dava Sobel, Longitude: The true story of a lone genius who solved the greatest scientific problem of his time, 1995, Penguin Books